\documentclass[conference]{IEEEtran}
\IEEEoverridecommandlockouts
\usepackage{cite}
\usepackage{amsmath,amssymb,amsfonts,latexsym}
\usepackage{amsthm}
\usepackage{braket}
\usepackage{graphicx}
\usepackage{textcomp}
\usepackage{xcolor}
\usepackage{mathtools}
\usepackage{multirow}
\usepackage{url}
\usepackage{booktabs}
\usepackage{hyperref}
\usepackage{lipsum}
\usepackage{nicematrix}
\usepackage[all]{xy}
\usepackage{tikz-cd}
\usepackage{tikz}
\usepackage{makecell}
\usepackage{caption}
\usepackage{subcaption}
\usepackage{quantikz}

\usetikzlibrary{calc}
\usetikzlibrary{intersections,shapes.arrows}
\usetikzlibrary{arrows.meta, positioning, backgrounds}

\usepgfmodule{nonlineartransformations}
\def\fluttertransform{%
    \pgfgetlastxy\x\y
    \pgfpoint{\x+sin(\y)}{\y+sin(\x)*(30-\x/2)+\x/10}
}

\def \s{\sigma}

\def \mcal{\mathcal}

\newtheorem{definition}{Definition}

\newtheorem*{theorem*}{Theorem}

\theoremstyle{definition}
\newtheorem{example*}{Example}



\def\BibTeX{{\rm B\kern-.05em{\sc i\kern-.025em b}\kern-.08em
    T\kern-.1667em\lower.7ex\hbox{E}\kern-.125emX}}

\makeatletter
\newcommand{\linebreakand}{%
  \end{@IEEEauthorhalign}
  \hfill\mbox{}\par
  \mbox{}\hfill\begin{@IEEEauthorhalign}
}
\makeatother

\begin{document}

\title{Adaptive Non-local Observable on Quantum Neural Networks\\

\thanks{
This work is supported by Laboratory Directed Research and Development Program \#24-061 of Brookhaven National Laboratory and National Quantum Information Science Research Centers, Co-design Center for Quantum Advantage (C2QA) under Contract No. DE-SC0012704.

The views expressed in this article are those of the authors and do not represent the views of Wells Fargo. This article is for informational purposes only. Nothing contained in this article should be construed as investment advice. Wells Fargo makes no express or implied warranties and expressly disclaims all legal, tax, and accounting implications related to this article.}
}



\author{
\IEEEauthorblockN{
    Hsin-Yi Lin$^1$\IEEEauthorrefmark{1},
    Huan-Hsin Tseng$^2$\IEEEauthorrefmark{2},
    Samuel Yen-Chi Chen$^3$\IEEEauthorrefmark{3}, Shinjae Yoo$^2$\IEEEauthorrefmark{4}
}
\IEEEauthorblockA{{}$^1$Department of Mathematics and Computer Science, Seton Hall University, South Orange, NJ, USA}
\IEEEauthorblockA{{}$^2$AI \& ML Department, Brookhaven National Laboratory, Upton, NY, USA}
\IEEEauthorblockA{{}$^3$Wells Fargo, New York, NY, USA}

\IEEEauthorblockA{Email:
\IEEEauthorrefmark{1} hsinyi.lin@shu.edu,
\IEEEauthorrefmark{2} htseng@bnl.gov,
\IEEEauthorrefmark{3} yen-chi.chen@wellsfargo.com,
\IEEEauthorrefmark{4} syjoo@bnl.gov}}

\maketitle


\begin{abstract}

Conventional Variational Quantum Circuits (VQCs) for Quantum Machine Learning typically rely on a fixed Hermitian observable, often built from Pauli operators. Inspired by the Heisenberg picture, we propose an \emph{adaptive non-local measurement} framework that substantially increases the model complexity of the quantum circuits. Our introduction of \emph{dynamical Hermitian observables} with evolving parameters shows that optimizing VQC rotations corresponds to tracing a trajectory in the observable space. This viewpoint reveals that standard VQCs are merely a special case of the Heisenberg representation.


Furthermore, we show that properly incorporating variational rotations with non-local observables enhances qubit interaction and information mixture, admitting flexible circuit designs. Two non-local measurement schemes are introduced, and numerical simulations on classification tasks confirm that our approach outperforms conventional VQCs, yielding a more powerful and resource-efficient approach as a Quantum Neural Network.





\end{abstract}

\begin{IEEEkeywords}
Variational Quantum Circuits, Quantum Machine Learning, Quantum Neural Networks, non-local observables, Heisenberg representations, Hermitian operators, classifications.
\end{IEEEkeywords}

\section{\label{sec:Indroduction}Introduction}
Quantum Machine Learning (QML) is a developing field that leverages the principles of quantum mechanics to advance machine learning (ML) models. With the rapid advancement of quantum computing hardware, QML aims to exploit quantum phenomena—such as superposition, entanglement, and quantum interference—to provide computational advantages over classical approaches. Despite the current limitations of quantum hardware, hybrid quantum-classical algorithms have been developed to harness the strengths of both computing paradigms, allowing near-term quantum devices to contribute meaningfully to real-world ML tasks.

One of the most promising classes of hybrid quantum-classical algorithms is Variational Quantum Algorithms (VQAs) \cite{mcclean2016theory, cerezo2021variational, bharti2022noisy, bonet2023performance}. These algorithms optimize quantum circuit parameters using classical optimization techniques, including gradient-based methods \cite{mitarai2018quantum, cerezo2021cost} and gradient-free approaches such as surrogate methods \cite{ostaszewski2021structure}, simultaneous perturbation stochastic approximation (SPSA) \cite{cade2020strategies}, evolutionary algorithms \cite{chen2022variationalQRL}, and trust region methods like COBYLA for constrained variational optimization.

VQA is a fundamental optimization framework for various quantum models. Variational Quantum Eigensolver (VQE) \cite{peruzzo2014variational, kandala2017hardware, grimsley2019adaptive} is one such example, leveraging the VQA approach to approximate the ground state energy of a given Hamiltonian, making it particularly useful for quantum chemistry and material science. Another example is the Quantum Neural Network (QNN), which also employs the VQA framework but with the goal of fitting data for different machine learning applications. QNNs have been successfully applied to diverse ML tasks and scenarios, including classification \cite{farhi2018classification, chen2021end, qi2023qtnvqc, chen2022quantumCNN}, time-series prediction \cite{chehimi2024FedQLSTM, chen2024QFWP}, data compression\cite{romero2017quantum}, reinforcement learning \cite{chen2020variational}, generative modeling \cite{zoufal2019quantum} and natural language processing \cite{li2023pqlm, yang2022bert, di2022dawn, stein2023applying}. The explainability of the QNN classifier was also investigated in \cite{lin2024quantumGRADCAM}.

Variational Quantum Circuits (VQCs), or parameterized quantum circuits
(PQCs) \cite{benedetti2019parameterized}, often serve as building blocks for QNN implementation.  In conventional QNN architectures, local measurements are applied, where multiple basis measurements are performed on single qubits. Specifically, VQCs utilize Pauli observables $(\sigma_x, \sigma_y, \sigma_z)$ to compute expectation values $\bra{\psi}\sigma_i  \ket{\psi}$.
These expectation values serve as outputs for classification, regression, or loss function calculations in hybrid quantum-classical models. 


However, the limited connectivity of qubits in near-term quantum hardware significantly restricts the function space of VQCs, reducing model expressivity. This leads to low model capacity, hampering the ability of QNNs to effectively fit complex data distributions. While most QNN implementations rely on single-qubit measurements, certain strategies have been explored to incorporate multi-qubit measurements to improve model performance. For example, instead of measuring individual qubits separately, parity measurements determine whether the number of qubits in the $\ket{1}$ state is even or odd. This method has been applied in error detection \cite{riste2017demonstration} and quantum classifiers \cite{jerbi2023quantum}.
Another example is the pooling measurements in Quantum Convolutional Neural Networks (QCNNs). QCNNs are inspired by classical CNNs and employ multi-qubit pooling operations to extract hierarchical features from quantum data \cite{cong2019quantum}. These operations enhance QNNs' ability to process large-scale structured data. While these approaches improve model expressivity, they remain constrained by fixed measurement operators. Recently, a learnable observable framework \cite{chen2025learning} has been introduced, demonstrating improved performance by optimizing not only the circuit parameters but also the measurement observables. To further expand QNN capacity, non-local measurements—where observables act on multiple qubits—offer a promising alternative.


In this paper, we propose an innovative variational non-local measurement framework that significantly enhances QNN expressivity. Instead of relying on single-qubit observables, we consider trainable multi-qubit Hermitian matrices as observables. These observables are optimized jointly with standard variational quantum rotation angles, allowing the model to dynamically adapt its measurement strategy during training.

Through simulations, we demonstrate that our learnable observable framework outperforms VQCs with single-qubit observables. This improvement is attributed to the increased flexibility in measurement space, which allows the model to better capture complex correlations in data. Our experimental results further validate the superior performance of this approach, highlighting the advantages of trainable non-local observables in enhancing QNN capacity.

In sum, our \textbf{key contributions} are as follows:


\begin{itemize}
    \item Investigate the quantum circuit prediction mechanism, revealing the relation between VQCs and the dynamical Hermitian observables.
    \item Introduce Adaptive Non‐Local Observable (ANO) framework and investigates the non-local effect from theoretical and empirical perspectives.
    \item Devise two practical measurement schemes (sliding $k$‐local and pairwise combinatorial strategies) to operate ANO and provide strong empirical performance with improved parameter efficiency on real-world classification tasks.
\end{itemize}

By enabling greater measurement flexibility, our method significantly enhances QNNs' ability to learn complex functions, paving the way for more powerful quantum machine learning models.


The rest of the paper is organized as follows. In Sec.~\ref{sec_related_Work}, we highlight relevant work in previous literature. The details of the proposed framework are described in Sec.~\ref{sec_method}. The general structure of VQC is reviewed in Sec.~\ref{subsec_VQC structure}, followed by the discussion of variational non-local quantum measurements in Sec.~\ref{subsec: Navigating Observable Space} and model complexity in Sec.~\ref{subsec_method_connectivity}. To validate the proposed method, the experimental setup and results are explained in Sec.~\ref{sec_exp_results}. Finally, the conclusions are summarized in Sec.~\ref{sec_conclusion}.

\section{Related work}\label{sec_related_Work}


Measurement design is central to observable estimation and model performance, yet it has often been treated as a fixed component. Several studies have investigated ways to enhance measurement efficiency. One approach is to consider locally randomized measurements \cite{elben2019statistical, elben2020cross,zhu2021cross, haug2023quantum}. These works introduce local randomness into the measurement process, typically through random single-qubit unitaries, to extract statistical information with scalable protocols for efficient estimations of quantum systems.

Other than randomized strategies, there is also growing interest in adaptive and learnable measurement strategies as a pathway toward more efficient and accurate quantum inference. For instance, \cite{malmi2024enhanced} utilizes fixed and overcomplete POVMs (e.g., randomized Pauli measurements) to enhance observable estimation accuracy by learning optimized dual estimators that minimize variance. A more explicit form of learnable measurement is introduced in \cite{garcia2021learning}, where POVMs are parameterized as linear combinations of Pauli operators and jointly optimized with the quantum circuit via gradient descent. The measurement basis in this case is actively adapted during training to reduce estimation variance. Similarly, \cite{gebhart2023learning} reconstructs fixed but unknown POVMs through classical post-processing and explores Bayesian adaptive measurement strategies that dynamically select measurements from a predefined set (e.g., Pauli operators) to maximize information gain.



More recently, the adaptive approach is employed for QML in \cite{chen2025learning}, where general single-qubit measurements were considered. Instead of considering Pauli observables as previous works, they apply parameterized Hermitian observables, which are co-trained with the variational rotation gates. The observable itself is adaptive and learned in a differentiable manner, enhancing the model beyond what fixed Pauli measurements can offer for machine learning tasks.

Motivated by these developments, this work proposes adaptive non-local measurements for QML, aimed at substantially increasing the expressivity of QNN while maintaining resource efficiency. In contrast to prior approaches that restrict measurements to single-qubit observables or fixed local POVMs, we introduce parameterized multi-qubit observables as part of the QNN architecture. This design allows the model to capture non-local quantum correlations essential for representing twisted decision boundaries and learning complex data structures.


\section{VQC with Trainable Non-local Hermitians}\label{sec_method}


\subsection{General VQC Structure}\label{subsec_VQC structure}

Quantum computing utilizes the foundation of spinor dynamics. A qubit is a spin-$1/2$ particle described by a spin state (wavefunction) $\ket{\psi} = c_0 \ket{0} + c_1 \ket{1}$ with $c_0, c_1 \in \mathbb{C}$ such that the 1-qubit Hilbert space is $\mathcal{H}^1 \cong \mathbb{C}^2$ with the standard Hermitian metric $\langle \cdot, \cdot \rangle_{\mathbb{C}^2}$. Multi-qubits are described by multi-particle states of tensor products,
\begin{equation}
    \ket{\psi} = \sum_{i_1, \ldots, i_n} c_{i_1, \ldots, i_n} \, \ket{i_1} \otimes \cdots \otimes \ket{i_n}
\end{equation}
for $i_1, \ldots, i_n \in \{ 0, 1\}$ with $c_{i_1, \ldots, i_n} \in \mathbb{C}$. Vectors $ \ket{i_1, \cdots, i_n} := \ket{i_1} \otimes \cdots \otimes \ket{i_n}$ form an orthonormal basis for the $n$-qubit Hilbert space $\mathcal{H}^n \cong \mathbb{C}^{2^n}$. A unitary operator $U$ (also called a quantum \textit{gate}) then transforms from one state $\ket{\psi} \in \mcal{H}^n$ to another preserving the norm, $\langle U \psi | U \psi \rangle_{\mathbb{C}^{2^n}}  = \langle \psi | \psi \rangle_{\mathbb{C}^{2^n}} = 1$. We denote the collection of all gates by $\mcal{U}(\mathcal{H}^n) = \{ U: \mathcal{H}^n \to \mathcal{H}^n \, | \, U U^{\dagger} = U^{\dagger} U = I_{2^n} \}$.

A VQC is an approach to realize QML by seeking proper unitary transformations in $\mcal{U}(\mathcal{H}^n)$ to fit and learn data. 

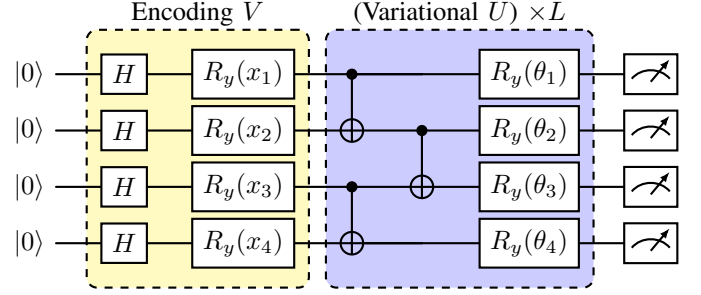
\begin{figure}[htbp]
 \vskip -0.1in
  \centering
    \begin{quantikz}[row sep=0.1cm, column sep=0.6cm]
        \lstick{$\ket{0}$} &\gate{H} \gategroup[wires=4,steps=2,style={dashed,rounded corners,fill=yellow!30,inner xsep=2pt},background]{Encoding $V$} & \gate{R_y(x_1)} & \ctrl{1} \gategroup[wires=4,steps=3,style={dashed,rounded corners,fill=blue!20,inner xsep=2pt},background]{(Variational $U$) $\times L$} & \qw &\gate{R_y(\theta_1)}  & \meter{} \\
        \lstick{$\ket{0}$} &\gate{H} & \gate{R_y(x_2)} & \targ{}  & \ctrl{1} &\gate{R_y(\theta_2)}  & \meter{} \\
        \lstick{$\ket{0}$} &\gate{H} & \gate{R_y(x_3)} & \ctrl{1} & \targ{}  &\gate{R_y(\theta_3)}  & \meter{} \\
        \lstick{$\ket{0}$} &\gate{H}  & \gate{R_y(x_4)} & \targ{}  & \qw      &\gate{R_y(\theta_4)}  & \meter{}
    \end{quantikz}
  \caption{A VQC diagrm of (\ref{E: encoding V}), (\ref{E: variational U}), which is also implemented in Sec.~\ref{sec_exp_results}. The variational block represented by a blue box is repeated $L$ times to increase the circuit depth.}
  \label{fig: Variational circuits}
\end{figure}

Specifically, let a classical dataset be $\mathcal{D} = \{(x^{(j)}, y^{(j)}) \, | \, x^{(j)} \in \mathbb{R}^n, y^{(j)} \in \mathbb{R}, \, j \in \mathbb{N} \}$ where $x^{(j)}$ denotes an input of sample $j$ and $y^{(j)}$ as the corresponding label. The VQC entails three steps to fit $\mathcal{D}$ (Fig.~\ref{fig: Variational circuits}): 

\begin{enumerate}
    \item \textbf{Encoding:} mapping a classical data $x \in \mathbb{R}^n$ (ignoring sample index $j$ for simplicity) to $V(x) \in \mcal{U}(\mcal{H}^n)$ by a choice of $V: \mathbb{R}^n \to \mcal{U}(\mcal{H}^n)$ and a random initial $\ket{\psi_0} \in \mcal{H}^n$. Then
        \begin{equation}\label{E: encoding}
            \ket{\psi_{x}} := V(x)\ket{\psi_0}
        \end{equation}
        embedded with the information of $x$ is called an \textit{encoded} state.
    \item \textbf{Variation:} A unitary $U(\theta) \in \mathcal{U}(\mathcal{H}^n)$ of tunable parameters $\theta$ is built to further transform $\ket{\psi_x}$ in (\ref{E: encoding}). Freely adjusting parameters $\theta$ allows $U(\theta) \ket{\psi_{x}}$ to navigate different states in $\mcal{H}^n$ as much as possible.
    \item \textbf{Measurement:} A constant Hermitian operator $H$ is chosen such that $\bra{\psi} H \ket{\psi}$ yields a real number, viewed as a VQC prediction to approach a designated target $y$.
\end{enumerate}

Typical constructions of quantum circuits include combinations of Hadamard gate $\texttt{H}$, CNOT gate $\mcal{C}$, and gates generated by Pauli matrices $\mathcal{P} = \{I, \s_1, \s_2, \s_3 \}$,
$\phi \mapsto \{e^{ -\frac{i}{2} \phi \s_1}, e^{ -\frac{i}{2} \phi \s_2}, e^{ -\frac{i}{2} \phi \s_3}\}$, also called \textit{rotation gates} denoted by $\{ R_x(\phi), R_y(\phi), R_z(\phi) \}$, respectively. For example, an $n$-qubit \textit{encoding} of input $x = (x_1, \ldots, x_n) \in \mathbb{R}^n$ can be,
\begin{equation}\label{E: encoding V}
    V(x) = \bigotimes_{q=1}^n \left(e^{ -\frac{i}{2} x_q \, \s_{k_q} } \circ \texttt{H}\right)
\end{equation}
with Pauli matrix index $k_q \in \{0, 1, 2, 3\}$ of $| \mathcal{P} |$. Similarly, a multi-qubit variational gate can be constructed as follows,
\begin{equation}\label{E: variational U}
    U(\theta) = \prod_{\ell=1}^L  \left( \bigotimes_{q=1}^n e^{ -\frac{i}{2} \theta_q^{(\ell)} \s_q  } \right) \circ  \mathcal{C}_{\ell} \in \mcal{U}(\mcal{H}^n)
\end{equation}
with tunable parameters $\theta = (\theta_1^{(1)} , \ldots, \theta_n^{(L)}) \in \mathbb{R}^{n \times L}$ (Fig.~\ref{fig: Variational circuits}).

The three steps above together yield a VQC function,
\begin{equation}\label{E: VQC output}
    \langle H \rangle (x; \theta) := \bra{\psi_0} V^{\dagger}(x) \underbracket{U^{\dagger}(\theta) \, H \, U(\theta)} V(x) \ket{\psi_0}
\end{equation}
with a loss function to guide the search of parameters $\theta$ under $\mcal{D}$ such as,
\begin{equation}\label{E: loss}
L(\theta; \mathcal{D}) = \frac{1}{| \mcal{D} |} \sum_{j=1}^{| \mcal{D} |} \| \braket{H} (x^{(j)}; \theta) - y^{(j)} \| ^2
\end{equation}

It is important to note that the variational $U(\theta)$ is the \textit{only} learnable component in the VQC, while the encoding map $V$ and the measurement Hermitian $H$ remain fixed. Thus, the goal of the VQC is to find a proper $U(\theta)$ minimizing the prediction loss \eqref{E: loss}.

Hermitian measurements, also referred to as \textbf{observables}, carry important meanings in physics, such as energy distributions of quantum systems. However, the significance of their roles in VQCs is seldom discussed when the standard choice is a fixed Pauli matrix in $\mathcal{P}$. Motivated by the Heisenberg and the Interaction pictures in Quantum Mechanics, we investigate alternative viewpoints on how these observables contribute to the VQC's prediction mechanism.

One inspiration for this proposed method is the Heisenberg representation and the interaction picture. Recall that the Heisenberg picture~\cite{tannoudji2020quantum} provides a distinct view as opposed to Schr\"{o}dinger’s in the early development of Quantum Mechanics. Heisenberg’s interpretation was that quantum evolution is characterized by \textit{dynamical observables} (Hermitian operators), while quantum states remain static. Current VQC majorly adopts Schr\"{o}dinger's perspective, where states evolve in time via unitary transformations, such as the movement in the Bloch sphere.

\subsection{Adaptive Non-local Observables}\label{subsec_method_nonlocal}

 We propose adaptive non-local observables (ANO) as a generalization to the conventional VQC by considering the following form of the \textbf{$k$-local observable}. 
 
 Consider an n-qubit system and $k\leq n$. Let $K=2^k$, an adaptive k-local observable 
\begin{equation}\label{E: non-local Hermitian}
  H(\phi) =   \begin{pmatrix}
c_{11} & a_{12} + i b_{12} & a_{13} + i b_{13} & \cdots & a_{1K} + i b_{1K}  \\
* & c_{22}  & a_{23} + i b_{23}  & \cdots & a_{2K} + i b_{2K}  \\
* & * & c_{33}  & \cdots & a_{3K} + i b_{3K}  \\
\vdots & \vdots & \vdots & \ddots & \vdots \\
* & * & * & \cdots & c_{KK}
\end{pmatrix}
\end{equation}
where $\phi = \left( a_{ij}, b_{ij}, c_{ii} \right)_{i, j=1}^K$ are arbitrary $K^2$ real numbers $\left( 2 \times \frac{K(K-1)}{2} + K \right)$ and the lower triangle entries are defined by the corresponding complex conjugates such that $H(\phi) = (h_{ij}) = (\overline{h_{ji}}) = H^{\dagger}(\phi)$.

\begin{figure}[htbp]
    \centering 
\begin{tikzpicture}[scale=1]
\begin{scope}[yshift=20mm]
\pgftransformnonlinear{\fluttertransform}
\draw [fill=red!15] plot [smooth cycle]
coordinates {(-1.14,-1)(-0.84, -.18) (-0.04, 0.3) (2.24, 0) %
(4.48, -0.56) (4.48, -1.46) (3.38,-1.84)(0.38, -1.28)};
\end{scope}

\coordinate (p1) at (0.1, 1.6);
\coordinate (p2) at (1.4, 1.9);
\coordinate (p3) at (2.5, 0.95);
\coordinate (p4) at (3.8, 1.0);

\filldraw (4.1, 2.1) node[left] {$\mathbb{R}^{N \times N}$};

\filldraw (p1) circle (1pt) node[above] {\footnotesize{$H(\theta_0)$}};
\filldraw (p2) circle (1pt) node[above] {\footnotesize{$H(\theta_1)$}};
\filldraw (p3) circle (1pt) node[below] {\footnotesize{$H(\theta_2)$}};
\filldraw (p4) circle (1.8pt) node[below] {\footnotesize{$H(\theta^*)$}};



\draw[solid] (p1) -- (p2) -- (p3) -- (p4);
\draw[dashed] (p1) .. controls ($(p1)!0.5!(p2)-(0,0.3)$) .. node[pos=0.5, below]{\footnotesize{$U$}} (p2);
\end{tikzpicture}
\caption{By (\ref{E: rotate Hermitian}), a VQC varying $U(\theta)$ is equivalent to searching for observable $H(\theta) := U^{\dagger}(\theta) H U(\theta)$ minimizing loss (\ref{E: loss}).}
\label{fig: search Hermitian}
\end{figure}
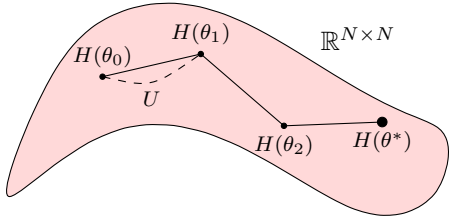

An important observation is that the variational rotation $U(\theta)$ in (\ref{E: VQC output}) rotates from one Hermitian $H$ to \textit{another}:
\begin{equation}\label{E: rotate Hermitian}
    H \mapsto U^{\dagger}(\theta) \, H \, U(\theta).
\end{equation}
Indeed, one sees that if we define $H(\theta) \overset{\text{def}}{=} U^{\dagger}(\theta) H U(\theta)$, it is also Hermitian whenever $U(\theta)$ is unitary. From the viewpoint of (\ref{E: rotate Hermitian}), it amounts to saying that the VQC is to find an \textit{optimal observable} $H(\theta)$ minimizing loss (\ref{E: loss}), Fig.~\ref{fig: search Hermitian}. Consequently, admitting direct search for Hermitian observables \textit{not only} includes the conventional VQC and also potentially allows one to omit the variational gate $U(\theta)$ optimization.

It is worth noticing that the typical 1-parameter subgroup parametrization method, such as $\theta \mapsto e^{\theta A}$ of a unitary group $\mcal{U}(\mcal{H}^1)$, does not directly translate to $\mathbb{H}(k)$. In fact, the set of all Hermitian operators on $k$-qubits, denoted as $\mathbb{H}(k) = \{H: \mcal{H}^k \to \mcal{H}^k \, | \, H = H^{\dagger}\}$, does not form a Lie group (not even a group with respect to the multiplication). Without the Lie structures (group, algebra), we opt to parameterize ANO by the formulation (\ref{E: non-local Hermitian}).

Note that the mapping $\phi \mapsto H(\phi)$ in (\ref{E: non-local Hermitian}) establishes an isomorphism from $\mathbb{R}^{K^2}$ to $\mathbb{H}(k)$, which indicates any observable has a unique correspondence $\phi$ and vice versa. (\ref{E: non-local Hermitian}) yields \textbf{$k$-local observables} in general due to the following relation.
\begin{definition}[Exact $k$-local operator]
Let $\mcal{L}(\mcal{H}^n)$ denotes all linear transformations on $\mcal{H}^n$. If $A \in \mcal{L}(\mcal{H}^n)$ is called an \textbf{exact $k$-local} operator ($k \leq n$) if it can be written as
\begin{equation}\label{E: exact n-local}
    A = I \otimes \cdots \otimes I \otimes Q \otimes I \otimes \cdots \otimes I,
\end{equation}
where $Q \in \mcal{L}(\mcal{H}^k)$ acts non-trivially on \textbf{exactly} $k$ of the $n$ qubits with $I$ denoting the identity on the remaining qubits.
\end{definition}


Indeed, for $Q \in \mcal{L}(\mcal{H}^k)$ in (\ref{E: exact n-local}), let a vector $v \in \mcal{H}^k$ be represented by a tensorial basis $\beta = \{ e_{i_1} \otimes \cdots \otimes e_{i_k} \}_{i_1, \ldots, i_k}$,
\begin{equation}
    v = \sum_{i_1, \ldots, i_k}^d v_{i_1, \ldots, i_k} \, e_{i_1} \otimes \cdots \otimes e_{i_k} 
\end{equation}
with $v_{i_1, \ldots, i_k} \in \mathbb{C}$ and $\{ e_1, e_2, \ldots, e_d \}$ a basis of $\mcal{H}^1$. Then,
\begin{equation}\label{E: Hv}
\begin{aligned}
    Q v &= \sum_{i_1, \ldots, i_k} v_{i_1, \ldots, i_k} \, Q( e_{i_1} \otimes \cdots \otimes e_{i_k} ) \\
           &= \sum_{i_1, \ldots, i_k}^d \sum_{j_1, \ldots, j_k}^d Q_{i_1, \ldots, i_k}^{j_1, \ldots, j_k} \, v_{i_1, \ldots, i_k} \, e_{j_1} \otimes \cdots \otimes e_{j_k}
\end{aligned}
\end{equation}
where $[Q]_{\beta} = \left( Q_{i_1, \ldots, i_k}^{j_1, \ldots, j_k} \right) \in \mathbb{C}^{d^k \times d^k}$ is the matrix representation of $\beta$. Then (\ref{E: Hv}) can be written in $\beta$ as
\begin{equation}
    \left[ Qv \right]_{\beta} = \left[ Q \right]_{\beta} \left[ v\right]_{\beta} = \begin{pmatrix}
  & \vdots &   \\
\cdots & Q_{i_1, \ldots, i_k}^{j_1, \ldots, j_k} & \cdots \\
  & \vdots &   
\end{pmatrix}
\begin{pmatrix}
v_1\\
v_2\\
\vdots\\
v_{d^k}
\end{pmatrix}
\end{equation}
If $Q$ is Hermitian and non-trivial, the matrix elements are written in (\ref{E: non-local Hermitian}). Thus, (\ref{E: non-local Hermitian}) represents an $n$-local observable in general.

To connect with the previous literature, we note that our definition of the exact $k$-local operator is strictly contained in the usual definition of $k$-local operators~\cite{kitaev2002classical}, which includes all exact $r$-locals of $r \leq k$. For instance, a 2-local Hamiltonian of spin glass~\cite{oliveira2005complexity},
\[
H = \sum_{i,j \in E} J_{ij} \, \sigma_z^{(i)} \otimes \sigma_z^{(j)} + \sum_{i \in V} c_i \, \sigma_z^{(i)}
\]
contains \textit{exact} 2-locals of the form $ \sigma_z^{(i)} \otimes \sigma_z^{(j)} $ at site $(i,j)$ and 1-locals $\sigma_z^{(i)}$.

We also remark that the prior work~\cite{chen2025learning} considered the special case of \textbf{1-local} observables,
\begin{equation}\label{E: 1-local Hamiltonian}
    H_1 \otimes I \otimes \cdots \otimes I, \quad I \otimes H_2 \otimes  \cdots \otimes I, \, \, \ldots  \quad I \otimes  \cdots \otimes H_n,
\end{equation}
with $H_i \in \mcal{L}(\mcal{H}^1)$ and have shown that learning to measure 1-local observables outperforms the conventional VQC, even as (\ref{E: 1-local Hamiltonian}) are not sufficient to represent arbitrary measurements yet.

\subsection{Navigating Observable Space via ANO}\label{subsec: Navigating Observable Space}

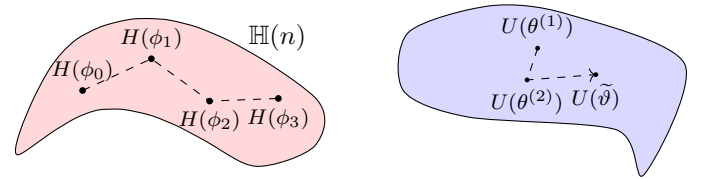
\begin{figure}[htbp]
    \centering 
\begin{tikzpicture}[scale=0.7]
    \begin{scope}[yshift=20mm]
    \pgftransformnonlinear{\fluttertransform}
    \draw [fill=red!15] plot [smooth cycle]
    coordinates {(-1.14,-1)(-0.84, -.18) (-0.04, 0.3) (2.24, 0) %
    (4.48, -0.56) (4.48, -1.46) (3.38,-2.3)(2.38,-2.1)(0.38, -1.28)};
    \end{scope}
    
    \coordinate (p1) at (0.1, 1.3);
    \coordinate (p2) at (1.4, 1.9);
    \coordinate (p3) at (2.5, 1.1);
    \coordinate (p4) at (3.8, 1.15);
    
    \filldraw (4.5, 2.3) node[left] {$\mathbb{H}(n)$};
    
    \filldraw (p1) circle (1.5pt) node[above] {\footnotesize{$H(\phi_0)$}};
    \filldraw (p2) circle (1.5pt) node[above] {\footnotesize{$H(\phi_1)$}};
    \filldraw (p3) circle (1.5pt) node[below] {\footnotesize{$H(\phi_2)$}};
    \filldraw (p4) circle (1.5pt) node[below] {\footnotesize{$H(\phi_3)$}};
    
    \draw[-, dashed] (p1) -- (p2) -- (p3) -- (p4);


    \begin{scope}[xshift=7cm]
    \begin{scope}[yshift=20mm]
    \draw [fill=blue!15] plot [smooth cycle]
    coordinates {(-0.84,-0.7)(-0.74, 0.4) (0.8, 0.9) (2.24, 0.75) %
    (4.48, -0.2) (3.7, -2.3) (3.18,-1.44)(0.38, -1.2)};

    \end{scope}

    \coordinate (p2) at (1.7, 2.1);
    \coordinate (p3) at (1.5, 1.5);
    \coordinate (p4) at (2.8, 1.6);
    
    
    \filldraw (p2) circle (1pt) node[above] {\footnotesize{$U(\theta^{(1)})$}};
    \filldraw (p3) circle (1pt) node[below] {\footnotesize{$U(\theta^{(2)})$}};
    \filldraw (p4) circle (1pt) node[below] {\footnotesize{$U(\widetilde{\vartheta})$}};
    
    \draw[->, dashed] (p2) -- (p3) -- (p4);
  \end{scope}
\end{tikzpicture}
\caption{[Left] Transitions in observable space $\mathbb{H}(n)$ when parametrizing from $\phi_1 \to \phi_2$ via (\ref{E: non-local Hermitian}). [Right] The corresponding transition in $\mcal{U}(\mcal{H}^n)$ via (\ref{E: Hermitian similarity 0}).}
\label{fig: Unitary param transition}
\end{figure}

Given an $n$-qubit system, the parameterization (\ref{E: non-local Hermitian}) allows us to explore the space $\mathbb{H}(n)$. A natural question to ask is: for any two Hermitian $H_1, H_2 \in \mathbb{H}(n) $,
\begin{equation}\label{E: Hermitian similarity 0}
    \text{Is there a  } \, U \in \mcal{U}(\mcal{H}^n) \quad \text{such that  } H_2 =  U^{\dagger} \, H_1 \, U?
\end{equation}
This can be viewed as an inverse operation of (\ref{E: rotate Hermitian}). If this is possible, the trajectories of adaptive Hermitians in ANO would be equivalent to the variational unitary operators in VQC via this 1-1 correspondence, see Fig.~\ref{fig: Unitary param transition}.

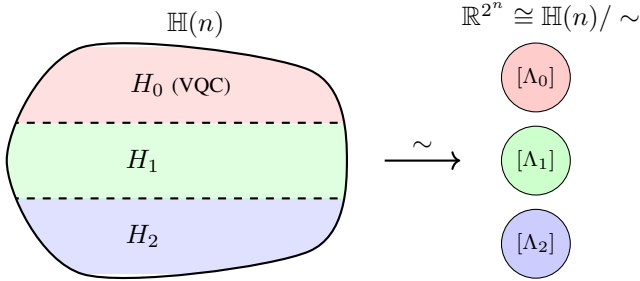
\begin{figure}[htbp]
    \centering 
\begin{tikzpicture}[scale=1]
  \def\Mcurve{
    plot [smooth cycle] coordinates {(0,0) (3,0.3) (3.5,1.5) (3,2.7) (0,3) (-1,1.5)}
  }
  
  \begin{scope}
    \clip \Mcurve;
    \fill[blue!20, opacity=0.6] (-2,0) rectangle (5,1);
  \end{scope}
  \begin{scope}
    \clip \Mcurve;
    \fill[green!20, opacity=0.6] (-2,1) rectangle (5,2);
  \end{scope}
  \begin{scope}
    \clip \Mcurve;
    \fill[red!20, opacity=0.6] (-2,2) rectangle (5,3);
  \end{scope}

  \begin{scope}
    \clip \Mcurve;
    \draw[dashed, thick] (-2,1) -- (5,1);
    \draw[dashed, thick] (-2,2) -- (5,2);
  \end{scope}

  \draw[thick] \Mcurve;
  \node at (1.5,3.3) {$\mathbb{H}(n)$};

  \node at (1.3,2.5) {$H_0$\footnotesize{ (VQC)}};
  \node at (0.8,1.5) {$H_1$};
  \node at (0.8,0.5) {$H_2$};

  \draw[->, thick] (4,1.5) -- (5,1.5) node[midway, above] {$\sim$};

  \node[draw, circle, fill=red!20, minimum size=3mm] (q3) at (6,2.6) {\footnotesize{$[\Lambda_0]$}};
  \node[draw, circle, fill=green!20, minimum size=3mm] (q2) at (6,1.5) {\footnotesize{$[\Lambda_1]$}};
  \node[draw, circle, fill=blue!20, minimum size=3mm] (q1) at (6,0.4) {\footnotesize{$[\Lambda_2]$}};
  \node at (6.2, 3.4) {$\mathbb{R}^{2^n} \cong \mathbb{H}(n) / \sim $};
\end{tikzpicture}
\caption{Observable space $\mathbb{H}(n)$ is partitioned into equivalent classes labeled by $\mathbb{R}^{2^n}$. In other words, $\mathbb{H}(n)$ is classified by its eigenvalues, and VQC falls in only one particular class.}
\label{fig: Hermitian equivalent classes}
\end{figure}

However, (\ref{E: Hermitian similarity 0}) is \textit{not} always true. The fact that (\ref{E: Hermitian similarity 0}) does not always hold indicates our proposed ANO contains a wider function class than the conventional VQC. This claim follows from the following result,
\begin{definition}[Unitary Similarity]
        Let $H_1, H_2 \in \mathbb{H}(n)$. If there exists $U \in \mcal{U}(\mcal{H}^n)$ such that $H_1 = U^{\dagger} \, H_2 \, U$ then $H_1 $ and $ H_2$ are said to be \textbf{unitarily similar}, denoted by $H_1 \sim H_2$.
\end{definition}
Immediately, we recognize the unitary similarity forms an \textit{equivalence relation} (reflexive, symmetric, and transitive). Since any Hermitian matrix is diagonalizable~\cite{horn2012matrix}, let $H \in \mathbb{H}(n)$ and write $H = U^{\dagger} \Lambda U$ for some unitary $U$ with $\Lambda = \text{diag}(\lambda_1, \ldots, \lambda_{2^n})$. As the eigenvalues of a Hermitian matrix must be \textit{real}, then $H \sim \Lambda \cong \mathbb{R}^{2^n}$ which implies the equivalence classes $\mathbb{H}(n) / \sim$ can be represented by $\mathbb{R}^{2^n}$. This leads us to the following conclusion.
\begin{theorem*}\label{Thm: Hermitian equivalence classes}
The unitary similarity defines an equivalence relation on $\mathbb{H}(n)$ and the resulting quotient space (collection of equivalence classes) $\mathbb{H}(n) / \sim$ can be identified with $\mathbb{R}^{2^n}$.
\end{theorem*}
This tells us that the unitary similarity partitions the observable space $\mathbb{H}(n)$ into categories of $\mathbb{R}^{2^n}$, see Fig.~\ref{fig: Hermitian equivalent classes}. Note that the theorem can be regarded as a variant of Sylvester Theorem~\cite{horn2012matrix}.

Looking back, now we see when (\ref{E: Hermitian similarity 0}) holds: it is precisely when two Hermitians share the \textit{same eigenvalues}. Consequently, the conventional VQC is a \textbf{special case} of ANO as it operates within only one single equivalence class, Fig.~\ref{fig: Hermitian equivalent classes}. Indeed, the conventional VQC selects and fixes an observable $H_0$ to perform rotations $U(\theta)$ by (\ref{E: rotate Hermitian}) to minimize loss (\ref{E: loss}). By the theorem, VQC is thereby confined in the equivalence class $[H_0]$, making it a restricted case of ANO.

Furthermore, Rayleigh Quotient~\cite{horn2012matrix} depicts that the VQC output (\ref{E: VQC output}) is bounded by the eigenvalues of an observable $H$,
\begin{equation}\label{E: Rayleigh Quotient}
    \lambda_{\text{min}} \leq \langle \psi , H \psi\rangle \leq \lambda_{\text{Max}} 
\end{equation}
with normalized state $\| \psi \| = 1$ and $\lambda_{\text{min}} \leq \cdots \leq \lambda_{\text{Max}}$ eigenvalues of $H$. 

Therefore, VQCs with fixed Pauli observables $\mathcal{P}$ will be further restricted in the equivalent class of $\lambda_{\text{min}} = -1$ and $\lambda_{\text{Max}} = +1$ to limit the prediction range $-1 \le \langle \psi, H \psi\rangle \le 1$, making it impossible to learn targets (labels) $y > 1$ and thus leaving loss~(\ref{E: loss}) irreducible.  While manual normalization on data is usually a common remedy, there exist occasions where such an operation is not suitable, such as when VQC serves as a middle layer or an encoder of an Auto-Encoder.

In contrast, ANO allows $H$ to have a broader spectrum, increases the circuit output range, and therefore improves the model capabilities. This ability to navigate different classes of equivalence (different eigenvalues) differentiates ANO from the conventional VQC in Sec.~\ref{subsec_VQC structure}.








\subsection{k-local ANO with Variational Rotations}\label{subsec_method_connectivity}



Theoretically, \textit{full} ANO ($k=n$) guarantees the coverage of all observable classes in $\mathbb{H}(n)$ (Fig.~\ref{fig: Hermitian equivalent classes}). However, in practice, the number of parameters grows rapidly with the number of qubits, especially when high-dimensional features are involved. In addition to the cost, excessive parameters can increase the risk of overfitting, similar to those observed in \cite{allen2019convergence}. These considerations urge us to investigate further the cases of smaller $k$-local ($k < n$) measurements and explore efficient ways to alleviate the limitation of smaller qubit connectivity.





Our investigation reveals that incorporating variational rotations with non-local ANO would assist \textit{feature interactions}, which is essential for model expressivity. To demonstrate this effect of rotations on ANO, we calculate explicitly to compare the model formulation on a 2-qubit system with: (1) only 1-local ANO and (2) 1-local ANO joining variational rotations in the following example.

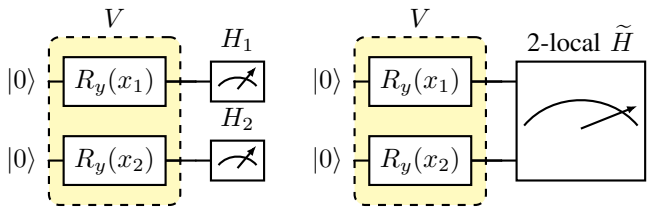
\begin{figure}[htbp]
 \vskip -0.2in
  \centering
  \begin{tikzpicture}[scale=0.8, node distance=1cm]
    \node (circuit1) {
    \begin{quantikz}[row sep=0.4cm, column sep=0.2cm]
        \lstick{$\ket{0}$} & \gate{R_y(x_1)} \gategroup[2,steps=1,style={dashed,rounded corners,fill=yellow!30,inner xsep=2pt},background]{$V$} & \qw & \qw  &\meter{H_1} \\
        \lstick{$\ket{0}$} & \gate{R_y(x_2)} & \qw & & \meter{H_2}
    \end{quantikz}
    };
    \node (circuit2) [right=0.15cm of circuit1] {
    \begin{quantikz}[row sep=0.4cm, column sep=0.2cm]
        \lstick{$\ket{0}$} & \gate{R_y(x_1)} \gategroup[2,steps=1,style={dashed,rounded corners,fill=yellow!30,inner xsep=2pt},background]{$V$} & \qw & \qw & \meter[2][2]{\text{2-local }\widetilde{H}} \\
        \lstick{$\ket{0}$} & \gate{R_y(x_2)} & \qw & \qw &
    \end{quantikz}
    };
  \end{tikzpicture}
  \caption{2-qubit circuits using \textit{only}: \textbf{[left]} 1-local, and \textbf{[right]} 2-local $\widetilde{H}$, without any entangling or variational rotations.}
  \label{fig: only local-measurements}
\end{figure}

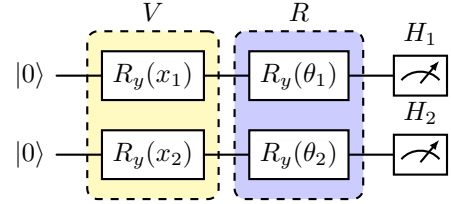
\begin{figure}[htbp]
\vskip -0.1in
  \centering
    \begin{quantikz}[row sep=0.4cm, column sep=0.6cm] 
        \lstick{$\ket{0}$} & \gate{R_y(x_1)} \gategroup[2,steps=1,style={dashed,rounded corners,fill=yellow!30,inner xsep=2pt},background]{$V$}  & \gate{R_y(\theta_1)} \gategroup[2,steps=1,style={dashed,rounded corners,fill=blue!20,inner xsep=2pt},background]{$R$} & \meter{H_1} \\
        \lstick{$\ket{0}$} & \gate{R_y(x_2)} & \gate{R_y(\theta_2)}  & \meter{H_2}
    \end{quantikz}
  \caption{1-local ANO + variational rotation $R$.}
  \label{fig: VNO + VQC}
\end{figure}

\begin{example*}[$k$-local ANO + rotations]\label{Example}
We use a simple 2-qubit system (Fig.~\ref{fig: only local-measurements},~\ref{fig: VNO + VQC}) to show the following model capacity relations.
\[
\text{1-local} \subseteq [\text{1-local + variational rotations} ]\subseteq \text{2-local}. \tag{$*$}
\]
Assume circuits in Fig.~\ref{fig: only local-measurements},~\ref{fig: VNO + VQC} have same encoding circuit $V$ and initial $\ket{\psi_0}$. Denote the encoded state as
\[
\ket{\psi} := V \ket{\psi_0} = (v_1, v_2, v_3, v_4) \in \mathbb{C}^4
\]
First, we compute the model outputs of 1-local measurements. Define the 1-qubit observables in Fig.~\ref{fig: only local-measurements} using (\ref{E: non-local Hermitian}),
\[
H_1 = \begin{pmatrix}
    h_{11}^{(1)} & h_{12}^{(1)} \\
    \overline{h}_{12}^{(1)} & h_{22}^{(1)}
\end{pmatrix}, \quad
H_2=\begin{pmatrix}
    h_{11}^{(2)} & h_{12}^{(2)} \\
    \overline{h}_{12}^{(2)} & h_{22}^{(2)}
\end{pmatrix}    
\]
with $h^{(i)}_{11}, h^{(i)}_{22} \in \mathbb{R}, h^{(i)}_{12} \in \mathbb{C}$; the corresponding 1-locals are,
\begin{equation}\label{E: 1-local obs}
\begin{aligned}
        \widetilde{H_1} &= H_1 \otimes I = \begin{pmatrix}
        h^{(1)}_{11} & 0 & h^{(1)}_{12} & 0 \\
        0 & h^{(1)}_{11} & 0 &  h^{(1)}_{12} \\
        \overline{h}_{12}^{(1)} & 0 & h^{(1)}_{22}& 0 \\
        0 & \overline{h}_{12}^{(1)} & 0 & h^{(1)}_{22}
        \end{pmatrix},\\
        \widetilde{H_2} &= I \otimes H_2 = 
        \begin{pNiceArray}{cc|cc}
          h^{(2)}_{11} & h^{(2)}_{12} & \Block{2-2}<\Large>{\mathbf{0}} \\
          \overline{h}_{12}^{(2)} & h^{(2)}_{22} \\
          \hline
          \Block{2-2}<\Large>{\mathbf{0}} && h^{(2)}_{11} & h^{(2)}_{12} \\
          &  &\overline{h}_{12}^{(2)} & h^{(2)}_{22}
        \end{pNiceArray}
\end{aligned}
\end{equation}
For simplification, we consider the special case of real encoded state $(v_1, v_2, v_3, v_4) \in \mathbb{R}^4$. In this case, the 1-local observables have the following expectation values as model outputs:
\begin{equation}\label{E: 1-local expectations}
\begin{aligned}
    \langle \widetilde{H_1} \rangle_{\psi} &= h^{(1)}_{11} \left(v_1^2 + v_2^2 \right) + 2\Re{h^{(1)}_{12}} \left( v_1 v_3 + v_2 v_4 \right) +  h^{(1)}_{22} \left( v_3^2 + v_4^2 \right)\\
    \langle \widetilde{H_2} \rangle_{\psi} &= h^{(2)}_{11} \left(v_1^2 + v_3^2 \right) + 2\Re{h^{(2)}_{12}} \left( v_1 v_2 + v_3 v_4 \right) +  h^{(2)}_{22} \left( v_2^2 + v_4^2 \right)
\end{aligned}
\end{equation}
Next, when the 1-local measurements are incorporated with a variational rotation $R:= R_y(\theta_1) \otimes R_y(\theta_2)$ of parameters $(\theta_1, \theta_2)$ (Fig.~\ref{fig: VNO + VQC}). The resulting expectation values are given by Equation (18) and (19).
\begin{figure*}[!h]
\normalsize
\setcounter{equation}{17}
\begin{multline}
    \langle R^{\dagger} \widetilde{H_1} R \rangle_{\psi} = h^{(1)}_{11} \left( (v_1^2 + v_2^2)\cos^2\frac{\theta_1}{2} + (v_3^2 + v_4^2) \sin^2\frac{\theta_1}{2} - ( v_1 v_3 + v_2 v_4)\sin\theta_1 \right)\\
        + \Re{h^{(1)}_{12}} \left( 2 \cos\theta_1 (v_1 v_3 + v_2 v_4) + (v_1^2 + v_2^2 - v_3^2 - v_4^2)\sin\theta_1 \right)\\
        + h^{(1)}_{22} \left( (v_1^2 + v_2^2)\sin^2\frac{\theta_1}{2} + (v_3^2 + v_4^2) \cos^2\frac{\theta_1}{2} + ( v_1 v_3 + v_2 v_4)\sin\theta_1 \right)
\end{multline}
\begin{multline}
    \langle R^{\dagger} \widetilde{H_2} R \rangle_{\psi} = h^{(2)}_{11} \left( (v_1^2 + v_3^2)\cos^2\frac{\theta_2}{2} + (v_2^2 + v_4^2) \sin^2\frac{\theta_2}{2} - ( v_1 v_2 + v_3 v_4) \sin\theta_2 \right)\\
           + \Re{h^{(2)}_{12}} \left( 2 \cos\theta_2 (v_1 v_2 + v_3 v_4) + (v_1^2 - v_2^2 + v_3^2 - v_4^2) \sin\theta_2 \right)\\
            + h^{(2)}_{22} \left( (v_1^2 + v_3^2)\sin^2\frac{\theta_2}{2} + (v_2^2 + v_4^2) \cos^2\frac{\theta_2}{2} + ( v_1 v_2 + v_3 v_4)\sin\theta_2 \right)
\end{multline}
\setcounter{equation}{20}
\hrulefill
\vspace*{4pt}
\end{figure*}
Comparing equations (17)-(19), we see the coupling between the observable elements $h_{jk}^{(i)}$ and feature vector $(v_1, \ldots, v_4)$ is more versatile in (18), (19) than in the 1-local case (\ref{E: 1-local expectations}). For instance, in (18), $h_{11}^{(1)}$ perceives all four features $v_1$ to $v_4$ at once, whereas in (\ref{E: 1-local expectations}) $h_{11}^{(1)}$, it only sees $v_1, v_2$, not knowing the existence of $v_3, v_4$. Thus, when optimizing variable $h_{jk}^{(i)}$, (18) provides more information from input, giving better guidance than the pure 1-local adaptation (\ref{E: 1-local expectations}). This can also be seen by turning off the rotation $\theta_1 = \theta_2 = 0$ in (18), (19).


From the comparison, we observe that introducing rotations enhances feature mixing within the quantum state, helping to compensate for the limited connectivity in lower $k$-local measurements. The above argument completes the illustration of the first subset relation in ($*$).

As discussed in Sec.~\ref{subsec: Navigating Observable Space}, the full ANO ($2$-local in this case) includes all possible Hermitian matrices that can be reached by applying rotations to 1-local observables, which therefore implies the second subset relation in ($*$). The trade-off, however, is to increase $k$-locality to substitute for variational rotations. $\blacksquare$
\end{example*}

The above heuristic shows how variational rotations increase feature interactions within the ANO, enhancing information mixing. This suggests that combining variational rotations with smaller $k$-local ANO can serve as an efficient alternative to the full adaptive observable ($k=n$). To validate this insight, we conduct experiments and introduce two local measurement configurations designed to improve feature interaction efficiency. The details are presented in the following section.

\section{Experiments}\label{sec_exp_results}

We evaluate the efficient $k$-local observables joining rotations in Sec.~\ref{subsec_method_connectivity} with two classification tasks: the \textbf{Banknote Authentication} dataset \cite{banknote_authentication_267} and \textbf{MNIST}\cite{lecun1998gradient}.

\subsection{Experimental setup}
\textbf{Datasets:} 1) The {Banknote Authentication} consists of 1,372 samples, each with 4-dimensional features extracted from wavelet-transformed images of banknotes: variance, skewness, curtosis, and entropy. The task is binary classification—determining whether a banknote is genuine or forged. The dataset is split $90/10$ into training and testing sets.

2) The {MNIST} comprises grayscale images of handwritten digits (0–9), each of size $28\times28$, for a 10-class classification task. Due to computational constraints, the images are resized to $4\times4$ via interpolation and use a subset of 10,000 samples. A random split of 9,000 training and 1,000 test images is performed for evaluation.


\vspace{1em} \noindent \textbf{Circuit structure:} The circuit structure before measurements in all experiments is shown in Fig.~\ref{fig: Variational circuits}. Unless specified otherwise, the same architecture is employed throughout.

An input \(x \in \mathbb{R}^n\) is encoded into an $n$-qubit system via (\ref{E: encoding V}) from an initial state $\ket{0}^{\otimes n}$. Next, the variational $U(\theta)$ in (\ref{E: variational U}) is applied, consisting of (a) a chain of CNOT gates between adjacent qubits in a linear topology $(1,2), (2,3), \dots, (n-1,n)$ and (b) a tensor product of 1-qubit rotations $R_y$ applied independently to each qubit with parameter $\theta$. This layer is repeated \(L=4\) times in all experiments.



\begin{figure}[!h]
 \vskip -0.0in
  \centering
  \begin{tikzpicture}[scale=0.8, node distance=1.5cm]
    \node (circuit1) {
    \begin{quantikz}[row sep=0.3cm, column sep=0.5cm]
    \lstick{1} &\meter{} \gategroup[2,steps=1,style={dashed,rounded corners,fill=red!30,inner xsep=2pt},background]{}\\
    \lstick{2}  &\meter{} \\
    \lstick{$\vdots$} \\
    \lstick{$n$} & \meter{}
    \end{quantikz}
    };
    \node (circuit2) [right=0.2cm of circuit1] {
    \begin{quantikz}[row sep=0.4cm, column sep=0.5cm]
        \lstick{1} &\meter{} \gategroup[1,steps=1,style={dashed,rounded corners,fill=red!30,inner xsep=2pt},background]{}\\
        \lstick{$\vdots$} \\
        \lstick{5}  &\meter{} \gategroup[1,steps=1,style={dashed,rounded corners,fill=red!30,inner xsep=2pt},background]{}\\
        \lstick{$\vdots$} \\
        \lstick{$n$} & \meter{}
    \end{quantikz}
    };
    \node (circuit3) [right=0.2cm of circuit2] {
    \begin{quantikz}[row sep=0.4cm, column sep=0.5cm]
        \lstick{$\vdots$} \\
        \lstick{3} &\meter{} \gategroup[1,steps=1,style={dashed,rounded corners,fill=red!30,inner xsep=2pt},background]{}\\
        \lstick{$\vdots$} \\
        \lstick{$n$} & \meter{} \gategroup[1,steps=1,style={dashed,rounded corners,fill=red!30,inner xsep=2pt},background]{}
    \end{quantikz}
    };
  \end{tikzpicture}
  \caption{\textbf{[Scheme 2]} Choose a subset $\mcal{S}\subseteq\{1,\dots,n\}$ and measure every possible pair in $\mcal{S}$ with a 2-local ANO to yield a $\binom{|\mcal{S}|}{2}$-dimensional output.}
  \label{fig: Pairwise measurement}
\end{figure}
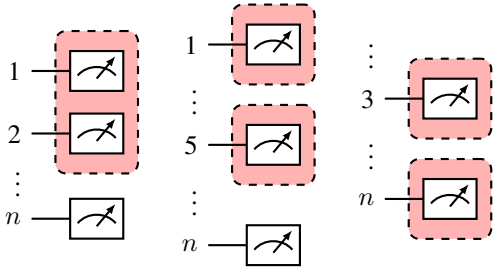

\vspace{1em} \noindent \textbf{Two Measurement Schemes:} The following experiments employ two distinct measurement schemes that differ in how qubits are grouped for measurement. See Fig.~\ref{fig: sliding k-local}, \ref{fig: Pairwise measurement} for illustrations. Qubit groupings are indicated by parentheses, e.g., $(2, 3)$ denotes a joint measurement on qubits 2 \& 3.

\vspace{1em} \noindent \subsubsection{\textbf{Sliding $k$-Local Measurements}}\label{subsubsec: sliding k-local}
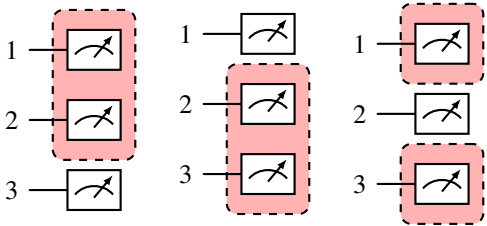
\begin{figure}[!h]
 \vskip -0.0in
  \centering
  \begin{tikzpicture}[scale=0.8, node distance=1.5cm]
    \node (circuit1) {
    \begin{quantikz}[row sep=0.4cm, column sep=0.5cm]
        \lstick{1}  &\meter{} \gategroup[2,steps=1,style={dashed,rounded corners,fill=red!30,inner xsep=2pt},background]{}\\
        \lstick{2} &\meter{} \\
        \lstick{3} &\meter{}
    \end{quantikz}
    };
    \node (circuit2) [right=0.2cm of circuit1] {
    \begin{quantikz}[row sep=0.4cm, column sep=0.5cm]
        \lstick{1}  &\meter{} \\
        \lstick{2}  &\meter{} \gategroup[2,steps=1,style={dashed,rounded corners,fill=red!30,inner xsep=2pt},background]{}\\
        \lstick{3}  &\meter{}
    \end{quantikz}
    };
    \node (circuit3) [right=0.2cm of circuit2] {
    \begin{quantikz}[row sep=0.4cm, column sep=0.5cm]
        \lstick{1}  &\meter{} \gategroup[1,steps=1,style={dashed,rounded corners,fill=red!30,inner xsep=2pt},background]{}\\
        \lstick{2} &\meter{} \\
        \lstick{3}  &\meter{} \gategroup[1,steps=1,style={dashed,rounded corners,fill=red!30,inner xsep=2pt},background]{}
    \end{quantikz}
    };
  \end{tikzpicture}
  \caption{\textbf{[Scheme 1]} Sliding $k$-local ANO with cyclic measurements. Here $k=2$, $n=3$.}
  \label{fig: sliding k-local}
\end{figure}

This scheme applies a sliding window of $k$-local measurement over an $n$-qubit system in a \textit{cyclic} fashion, forming $n$ overlapping $k$-qubit groups. Each group is measured with a distinct $k$-local ANO observable. For example, with $k=2$ and $n=3$, the measurement groups are:
\[
(1,2),\ (2,3),\ (3,1) \qquad \text{(see Fig.~\ref{fig: sliding k-local})}
\]
Each group, measured by a $k$-local ANO, produces a scalar output, yielding $n$ values in total. If the desired output dimension $d_{\text{out}} \leq n$, only the first $d_{\text{out}}$ outputs are retained. This approach provides structured, locality-aware measurements with efficient reuse across inputs. This approach is similar to the classical Convolution Neural Network (CNN) using finite horizon kernels (small $k$-local ANOs in our case) to patch local patterns into a global representation.

Although this cyclic sliding $k$-local ANO captures local structure, its parameter count grows exponentially with $k$. To further reduce parameters while considering various feature interactions, another 2-local measurement strategy can be conceived to evaluate all qubit pairs (or a selected subset) within a defined set as follows.

\vspace{1em} \noindent \subsubsection{\textbf{Pairwise Combinatorial Measurement}}\label{subsubsec: combinatorial pairwise measurement}
This scheme begins by selecting an observing qubit subset $\mcal{S}$ from the total $n$, $\mcal{S} \subseteq \{1, \dots, n\}$. For every distinct qubit pair $(i,j)$ within $\mathcal{S}$, a dedicated 2-local ANO measurement is performed, see Fig.~\ref{fig: Pairwise measurement}. For example, in a 6-qubit system:
\begin{itemize}
    \item If $\mathcal{S} = \{1,2,3,4,5,6\}$, then $\binom{6}{2} = 15$ pairs are measured.
    \item If $\mathcal{S} = \{1,3,5\}$, then 3 pairs: $(1,3),\ (1,5),\ (3,5)$ are measured.
\end{itemize}
The total number of pairwise measurements is $\binom{|\mathcal{S}|}{2}$, which defines the output dimension of this scheme. If this exceeds the desired output size, a linear layer is applied to project the output to the target dimension.

We emphasize that although each 2-local measurement captures only localized information, the combinatorial coverage within $\mathcal{S}$ allows these segments to collectively uncover global data structure.

This scheme is specifically designed to overcome the parameter-efficiency concern while retaining feature interaction capability. By solely relying on 2-local ($4\times4$ Hermitians) observables for each qubit pair, it tactically avoids the rapid parameter growth in higher $k$-local ANO while allowing pairwise feature mixtures.

\subsection{Results}

\subsubsection{\textbf{Banknote}}
Given the small feature dimension (4 features), we directly evaluate sliding $k$-local ANO of $k=1,2,3$ in (\ref{E: non-local Hermitian}) and compare them to a fixed Pauli measurement baseline. For each setup, we conduct ablation studies by removing rotational gates (``w/o Rs'' in Table~\ref{tab:result-banknote}) to assess their impact. Each experiment is repeated over 10 independent trials, with results summarized in Table~\ref{tab:result-banknote}.



First, we observe that even the simplest 1-local measurement significantly outperforms the fixed Pauli baseline. This conforms with the theoretical insights given in Sec.~\ref{subsec: Navigating Observable Space} that learnable Hermitian operators flexibly expand eigenvalues for local qubit groups. However, when removing the rotation gates, the performance drops as expected. As discussed in Example~\ref{Example}, this is because rotation gates enhance feature mixing and non-linear interactions so that the effects are particularly pronounced in the 1-local setting.




Interestingly, the effect of needing rotation gates diminishes as $k$ increases. In the 3-local case, performance remains essentially unchanged with or without the rotational layer.

In other words, as the non-local $k$ increases, the benefit of having rotations becomes less significant. For small $k$ (e.g., 1-local), rotation gates play a critical role in improving model expressivity—by mixing features and introducing non-linearities. However, when higher $k$-locals are involved, adaptive observables themselves already capture more complex, non-local correlations. As a result, the additional expressivity gained from including rotational gates becomes less crucial, and removing them has a smaller impact on performance.





Overall, performance improves with increasing $k$, benefiting from the richer structure of higher-order measurements. Nevertheless, due to the small input dimension, the 3-local slightly shows signs of overfitting, leading to a marginal drop in accuracy compared to the 2-local case. Thus, the best overall performance is attained by 2-local measurements with variational rotations.

\begin{figure}[htbp]
\vskip -0.15in
\begin{center}
\centerline{\includegraphics[width=1\columnwidth]{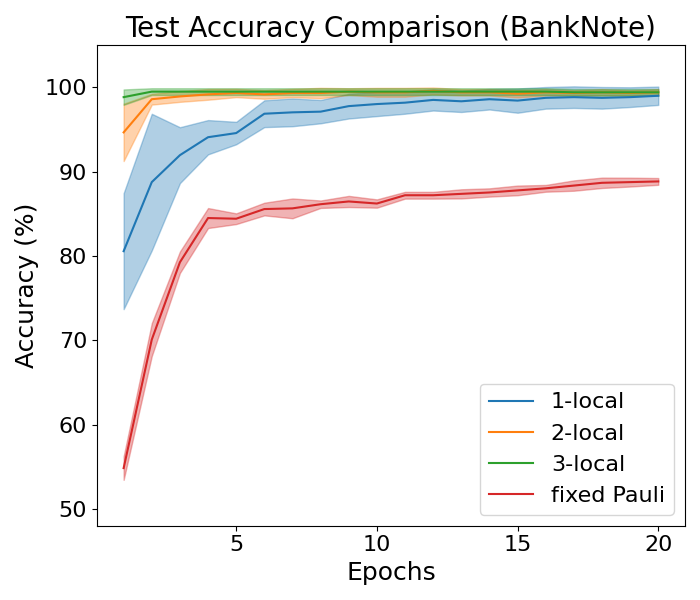}}
\caption{\textbf{Banknote test accuracy curves over epochs.} $k$-local models outperform the fixed Pauli baseline, with higher $k$ yielding faster convergence and better accuracy.
}
\label{fig: banknote}
\end{center}
\vskip -0.1in
\end{figure}

Fig.~\ref{fig: banknote} shows test accuracy over training epochs, highlighting the performance gap between the fixed Pauli baseline and the proposed $k$-local ANO. Notably, the 2-local and 3-local models converge rapidly with narrow confidence intervals, demonstrating stable learning ability.




\subsubsection{\textbf{MNIST}}

\begin{figure}[htbp]
\vskip -0.15in
\begin{center}
\centerline{\includegraphics[width=1\columnwidth]{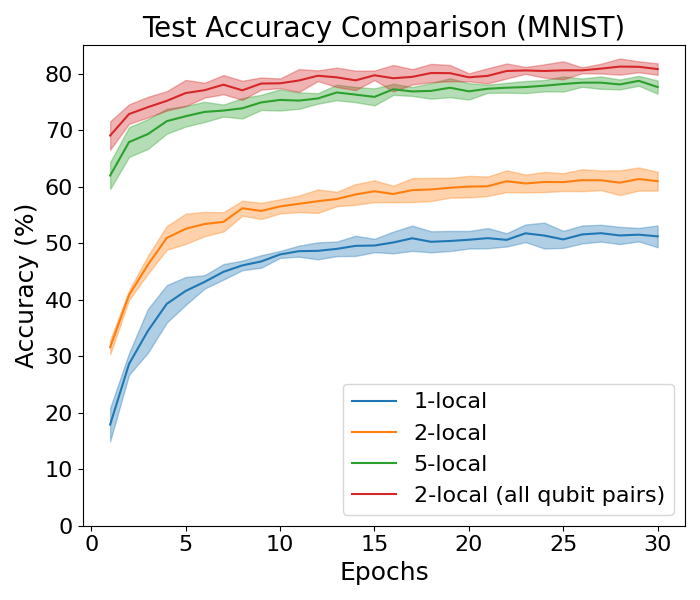}}
\caption{\textbf{MNIST test accuracy curves over epochs.} The pairwise 2-local model over all qubits (in red) consistently outperforms the best $k$-local methods.}
\label{fig: mnist}
\end{center}
\vskip -0.1in
\end{figure}

By resizing images to $4 \times 4$, we evaluate the two measurement schemes mentioned in Sec.~\ref{subsubsec: sliding k-local} and Sec.~\ref{subsubsec: combinatorial pairwise measurement}. Each setup is repeated over 5 trials. Results are summarized in Table~\ref{tab:main_mnist} and plotted in Fig.~\ref{fig: mnist}.



\vspace{1em} \noindent \textbf{Sliding $k$-local Measurement.}

In this scheme, we observe a consistent trend of accuracy improvement as $k$ increases, verifying the benefit of broader qubit interactions in measurement. However, the marginal gains (in test accuracy) diminish with each increment of $k$, Table~\ref{tab:main_mnist} (a):
\begin{itemize}
    \item 1-local $\to$ 2-local: $+9\%$
    \item 2-local $\to$ 3-local: $+8\%$
    \item 3-local $\to$ 4-local: $+6\%$
    \item 4-local $\to$ 5-local: $+4\%$.
\end{itemize}
Note the last observation from 4-local $\to$  5-local only improves 4\%. This diminishing return suggests a limit to the performance gains obtainable solely through increasing $k$.


However, learnable parameters grow quickly with $k$. Moving from 2-local $\to$ 5-local increases parameters by 46x (from $224 \to 10,304$), deriving only 17\% accuracy increase. This indicates a trade-off: while $k$-local measurements provide high model fitting capability, they tend to become redundant or inefficient when data is simple.





\vspace{1em} \noindent \textbf{Pairwise Combinatorial Measurement.}


We explore this scheme with three different cases of observing subset $\mcal{S}$, stated in Sec.~\ref{subsubsec: combinatorial pairwise measurement}:
\begin{itemize}
    \item $\mcal{S} = \{ \text{all 16 qubits} \}$,
    \item $\mcal{S} = \left\{1,3,5,7,9,11,13,15\right\}$ \quad (evenly-spaced 8 qubits)
    \item $\mcal{S} = \left\{1,4,7,10,13,16 \right\}$   \quad (sparse-distributed 6 qubits).
\end{itemize}

Notably, for the experiments in this scheme with MNIST, we \textit{completely} eliminate rotation gates to move beyond the traditional VQC framework.

Although this measurement scheme reduces the parameter numbers, performance remains competitive. Table~\ref{tab:main_mnist} (a) shows that a full 16-qubit ($4 \times 4$ images) ANO using pairwise measurement achieves the highest accuracy of 82\% with only 3,130 parameters, outperforming the 5-local model of 79.1\% accuracy using over three times as many parameters (10,304). 

Similarly, the 8-qubit ANO in this scheme achieves 74.9\% accuracy with just 738 parameters, comparable to the 4-local model’s 75.2\% but with more than 3x fewer parameters. These results demonstrate that pairwise measurement modeling, even in a limited 2-local form, can be highly effective when an observing set $\mcal{S}$ is suitably chosen. 

Collectively, these experiments reveal the potential of the proposed non-local ANO to be considered as an attempt beyond conventional VQC.

\begin{table}[t]
    \centering
    \caption{\textbf{[$k$-local Test Accuracy on Banknote]} $k$-local ANO outperform the fixed Pauli baseline. Rotation gates (``Rs'') boost performance, especially for lower $k$. For higher $k$, the impact of Rs diminishes.} 
    \resizebox{\columnwidth}{!}{%
    \begin{tabular}{p{0.12\linewidth} rr | cr}
        \toprule
        \textbf{} & \textbf{Param} \# &  \textbf{Acc (\%, w/ Rs)} & \textbf{Param \#} & \textbf{Acc (\%, w/o Rs)} \\
         \cmidrule(r){1-5}
        Pauli   & {16}                  & 89.0 $\pm$ 0.4                            & --              & {--} \\
        1-local & 24 (8+16)            & 99.3 $\pm$ 0.8                            & 8              & 73.3 $\pm$ 1.0 \\
        2-local & 48 (32+16)   & \textbf{99.7} $\mathbf{\pm}$ \textbf{0.4} & 32              &98.4 $\pm$ 0.3 \\
        3-local & 144 (128+16)          & 99.5 $\pm$ 0.4                            & 128    & \textbf{99.4} $\mathbf{\pm}$ \textbf{0.4} \\
        \bottomrule
\end{tabular}}
    \label{tab:result-banknote}
\end{table}

\begin{table}[ht]
\centering
    \caption{\textbf{[$k$-local ANO on MNIST]} Accuracy increases with larger $k$ but shows diminishing gains in the cases of sliding $k$-local ANO. Pairwise Combinatorial Measurement yields strong performance with fewer parameters and without the rotation gates, highlighting its efficiency in capturing feature interactions.} 
    \label{tab:mnist_fashionmnist}
  \centering
  \begin{subtable}[t]{0.45\textwidth}
    \centering
    \textbf{(a) Sliding \textit{k}-local}
        \vspace{2mm}
        \begin{tabular}{p{0.25\linewidth} r r}
            \toprule
            \textbf{} & \textbf{Param \#} & \textbf{Test accuracy (\%)} \\
            \cmidrule(r){1-3}
            {\makecell{1-local}} & {104} & 52.3 $\pm$ 1.9 \\
            {\makecell{2-local}} & {224} & 61.7 $\pm$ 2.2 \\
            {\makecell{3-local}} & {704} & 69.3 $\pm$ 1.0 \\
            {\makecell{4-local}} & {2624} & 75.2 $\pm$ 0.8 \\
            {\makecell{5-local}} & {10,304} & \textbf{79.1 $\pm$ 0.9} \\
        \bottomrule
        \end{tabular}
    \label{tab:subtable-a}
  \end{subtable}
  \hfill

  \begin{subtable}[t]{0.45\textwidth}
    \vspace{2mm}
    \centering
\textbf{(b) Pairwise Combinatorial}
        \vspace{2mm}
        \begin{tabular}{p{0.25\linewidth} r r}
            \toprule
            {} & \textbf{Param \#} & \textbf{Test accuracy (\%)} \\
            \cmidrule(r){1-3}
            {\makecell{6 qubits}} & {400} & {\makecell{62.4 $\pm$ 0.6}} \\
            {\makecell{8 qubits}} & {738} & {\makecell{74.9 $\pm$ 0.6}} \\
            {\makecell{16 qubits (all)}} & {3130} & {\makecell{\textbf{82.0 $\pm$ 1.1}}} \\
            \bottomrule
        \end{tabular}
    \label{tab:subtable-b}
  \end{subtable}
  \label{tab:main_mnist}
\end{table}

\section{Conclusion}\label{sec_conclusion}







In this work, we introduced an adaptive non-local measurement framework for QNNs that significantly enhances model expressivity by replacing traditional fixed Pauli matrices with trainable non-local Hermitian operators. 

Inspired by the Heisenberg representation, we consider dynamical Hermitian observables with evolving parameters (Sec.~\ref{subsec_method_nonlocal}), in contrast to the static observables of standard VQCs in the Schr\"{o}dinger picture (Sec.~\ref{subsec_VQC structure}). From this viewpoint, VQC and ANO appear to establish a one-to-one correspondence (Fig.~\ref{fig: Unitary param transition}), where every variational rotation $U(\theta)$ from VQC corresponds to a transition of an observable in the Hermitian space.


Investigating the quotient structure on Hermitian space by unitary similarity (Theorem, Sec.~\ref{subsec: Navigating Observable Space}) indicates that the conventional VQC is, in fact, only a special case of the ANO framework since the nature of a fixed observable necessarily restricts the VQC within only one equivalence class (Fig.~\ref{fig: Hermitian equivalent classes}). That is, the full ANO contains higher model complexity than the conventional VQC. While fully ANO makes it possible to eliminate variational rotation gates entirely, Sec.~\ref{subsec_method_connectivity} shows that strategically reintroducing rotations into smaller non-local observables can enhance qubit connectivity. This offers a parameter-efficient alternative that retains the expressive power of ANO.



To validate our theoretical framework, we conduct comprehensive simulations on two tasks, the Banknote Authentication and MNIST datasets. Due to the versatility of non-local ANO, we design two measurement schemes: sliding $k$-local measurements and pairwise combinatorial measurements. Results show that sliding significantly boosts classification performance, especially when paired with variational rotation gates. However, as the non-local size $k$ increases, the benefit of rotations diminishes, suggesting that the measurement observables alone can capture rich non-local correlations.

In contrast, pairwise combinatorial measurement offers a more parameter-efficient alternative. It reduces computational complexity and model size while maintaining, or even surpassing, the performance of more parameter-heavy $k$-local configurations.

Together, the ANO framework and these empirical studies reveal new directions for building expressive yet efficient QNNs, to provide varieties of quantum learning circuits.

\bibliographystyle{IEEEtran}
\bibliography{references,bib/qml_examples,bib/vqc,bib/explain_qml}

\end{document}